\documentclass[12pt]{iopart}
\usepackage{iopams}  
\usepackage{graphicx}
\usepackage{bm}
\usepackage{booktabs}
\usepackage{cite}

\def\buk{{\hat {\bm u}}_{\bm k}}
\def\bup{{\hat {\bm u}}_{\bm p}}
\def\buq{{\hat {\bm u}}_{\bm q}}

\def\ukp{u^+_{\bm k}}
\def\ukm{u^-_{\bm k}}
\def\bukp{{\bm u}^+_{\bm k}}
\def\bukm{{\bm u}^-_{\bm k}}
\def\bukpm{{\bm u}^\pm_{\bm k}}

\def\bk{{\bm k}}
\def\bp{{\bm p}}
\def\bN{{\bm N}}
\def\br{{\bm r}}
\def\bq{{\bm q}}
\def\bu{{\bm u}}
\def\bx{{\bm x}}
\def\bz{{\bm z}}
\def\bw{{\bm \omega}}
\def\hm{{\bm h}^-_{\bm k}}
\def\hp{{\bm h}^+_{\bm k}}
\def\hpm{{\bm h}^\pm_{\bm k}}

\usepackage{color}

\begin{document}

\title[]{Energy Cascade and Intermittency in Helically Decomposed Navier-Stokes Equations\footnote{Postprint version of the manuscript published in Fluid Dynamics Research {\bf 50}, 011420 (2018)}}

\author{Ganapati Sahoo$^{1,2}$\footnote{Corresponding 
author: ganapati.sahoo@gmail.com} and Luca Biferale$^2$\footnote{biferale@roma2.infn.it}}

\address{$^1$Department of Mathematics and Statistics and\\ Department of Physics, University of Helsinki, Finland.}
\address{$^2$Department of Physics and INFN, University of Rome Tor Vergata, Italy.}


\begin{abstract}
We study the nature of the triadic interactions in Fourier space for
three-dimensional Navier-Stokes equations based on the helicity content of the
participating modes. Using the tool of helical Fourier decomposition we are
able to access the effects of a group of triads on the energy cascade process
and on the small-scale intermittency.  We show that while triadic interactions
involving modes with only one sign of helicity results to an  inverse cascade
of energy and to a complete depletion of the intermittency, absence of such
triadic interactions has no visible effect on the energy cascade and on
the inertial-range intermittency of the three-dimensional Navier-Stokes
equations.

\end{abstract}

\vspace{2pc}

\maketitle

\section{Introduction \label{sec-1}} Triadic interactions among Fourier modes
are the fundamental building blocks of the Navier-Stokes equations and in
homogeneous and isotropic turbulent flow, the energy transfer is empirically
observed to be dominated by local interaction in Fourier space.  Inviscid
quadratic invariants of the Navier-Stokes equations are believed to be the key
to drive the direction and the fluctuations of the energy transfer such that in
three-dimensional (3D) turbulent flow there is a
forward energy cascade \cite{frisch,k41} from the injection scale down to the
smallest dissipative scale. In two-dimensions, the energy
cascades backward from small to the large scales because of the presence of two
sign-definite conserved quantities, energy and enstrophy
\cite{boffetta,kraichnan_2d}. Inverse energy cascades are also observed in
anisotropic 3D setups, e.g., in systems under strong rotation, with high shear,
under confinement along one direction and in conducting fluids \cite{mininni,deusebio,celani,brandenburg,lohse_arfm,sahoo2017prl,biferale2016prx} with
reduced dynamical equations \cite{embid1998,sukhatme2008} often used to highlight the underlying physical
processes in geophysical phenomena.  

It has also been argued that in three dimensions the second quadratic invariant
helicity plays an important role in the dynamics of the energy transfer
\cite{moffatt69,moffatt92,biferale2012,biferale2013,constantin,biferale-jstat,sahoo2015pre,sahoo2015epj,brissaud,laing,ditlevsen,stepanov,biskamp,holm,baerenzung,benzi,chen2003,chen2003prl}
even though it is not sign definite. Linear stability analysis of individual
triads (see Sec. 2 and \cite{waleffe}) shows that triads which couple Fourier
modes with the same helical content (homochiral triads) are capable of
transferring energy from small to large scale while all the other triads
(heterochiral) lead to a forward cascade. Indeed, in a direct numerical
simulation of a 3D homogeneous and isotropic turbulent flow an inverse energy
transfer is observed when Fourier modes with only one sign of helicity are kept
in the system \cite{biferale2012,biferale2013}.  

Earlier studies have shown that starting from an homochiral Navier-Stokes
simulation and by adding modes with the opposite sign of helicity, leads to a
transition from inverse to direct energy cascade
\cite{sahoo2015pre,herbert,kraichnan}. The transition is different, depending
on the protocol used to add heterochiral interactions
\cite{sahoo2015pre,sahoo2017prl}. Unfortunately, the only way to study all
potentially different triadic families is to recover to a fully spectral code
\cite{smith2005} with the consequential limitations in the computational
applications.  

This paper studies the dynamics of the three dimensional Navier Stokes
equations in the other limit: by restricting the evolution to heterochiral
interactions only. The aim is to understand how much the forward energy
transfer is affected by removing the homochiral triads, the ones that are
leading to an inverse energy transfer if taken alone. The problem is important
in connection with the presence of anomalous scaling and intermittency, i.e.,
the existence of strong non-Gaussian fluctuations in the inertial range of
turbulence \cite{frisch}. Indeed, it is not known how much intermittency
depends on the structure of the Fourier interactions.  

The paper is organized as follows. In Sec. \ref{sec-2} we introduce the helically
decomposed Navier-Stokes equations, in Sec. \ref{sec-3} we discuss the numerical
techniques used in our simulations, in Sec. \ref{sec-4} we show the results from our
direct numerical simulations followed by a discussion and conclusions in Sec.  \ref{sec-5}.

\section{Helically decomposed Navier-Stokes equations\label{sec-2}}

In a 3D periodic domain the velocity field can be expressed in Fourier series as
\begin{equation}
\bu(\bx) = \sum_{\bk} \buk e^{i\bk\cdot\bx}.
\end{equation}
 For low-Mach number flows the Fourier modes $\buk$ satisfy the incompressibility condition
\begin{equation}
\bk\cdot\buk=0 
\end{equation}
and therefore can be exactly decomposed in terms of the helically polarized waves as  \cite{constantin}:
\begin{equation}
\label{eq:dec}
  \buk  = \ukp \hp +\ukm \hm.
\end{equation}
 We write $\bukp  \equiv \ukp \hp$ and  $\bukm \equiv \ukm \hm$ so that
$\buk  = \bukp + \bukm$.
Here $\hpm$ are the eigenvectors of the curl operator such that 
\begin{equation}
i {\bk} \times \hpm = \pm k \hpm; 
\end{equation}
and are given by
\begin{equation}
\hpm = \hat{\nu}_{\bm k} \times
\hat{k} \pm i \hat{\nu}_{\bm k},
\end{equation}
where $ \hat{\nu}_{\bm k}$ is an
unit vector orthogonal to ${\bk}$ with the property $\hat{\nu}_{\bm k} = -
\hat{\nu}_{-\bk}$ and can be realized as
\begin{equation}
\hat{\nu}_{\bk} = \frac{{\bz} \times {\bk}}{ || {\bz} \times {\bk} ||},
\end{equation}
 for any arbitrary vector $\bz$.  The orthogonality conditions for the eigenvectors $\hpm$
are 
\begin{equation}
{\bf h}^{s}\cdot{\bf h}^{t\star}=2\delta_{st}, 
\end{equation}
where $s$ and $t$ are signs of the helicity which can be either $+$ or $-$ and $\star$ denotes the complex conjugate.
We can then define a projector 
\begin{equation}
  \label{eq:poperator}
  {\mathcal P}^\pm_{\bk} \equiv \frac {\hpm \otimes {\bf h}^{\pm\star}_\bk} {{\bf h}^{\pm\star}_\bk \cdot \hpm},
\end{equation}
which projects the Fourier modes of the velocity on eigenvectors $\hpm$
as
\begin{equation}
  \label{eq:projection}
  {\mathcal P}^\pm_{\bk} \buk = {\hat \bu}^\pm_\bk = u^\pm_\bk\hpm,
\end{equation}
We can then write the Navier-Stokes equations separately for velocities with positive or negative sign of helicity as:
\begin{equation}
\label{eq:NS}
\partial_t\bu^\pm(\bx) + {\cal D}^\pm \bN[\bu(\bx),\bu(\bx)] = -\nabla p(\bx) + \nu\Delta\bu^\pm(\bx),
\end{equation}
where ${\cal D}^{\pm}$ is the projector on $\hpm$, equivalent of ${\mathcal P}^{\pm}$, in real-space:
\begin{equation} 
  \label{eq:projector}
  {\cal D}^{\pm}{\bu}(\bx) \equiv \sum_{\bk} e^{i{\bm k}\bx}\,{\mathcal P}^{\pm}_{\bk} {\buk} = \bu^\pm(\bx).
\end{equation}
$\bN[\bu(\bx),\bu(\bx)]$ is the nonlinear term  of the Navier-Stokes equations which in
Fourier-space is given by
\begin{equation}
\hat \bN_\bk = -i \sum_{\bk+\bp+\bq=0} (\bq\cdot\bup)\buq,
  \label{eq:nlin}
\end{equation}
The inviscid invariants, the total energy and the total helicity, are sum of the contributions
from positively and negatively helical Fourier modes:
\begin{eqnarray}
    E &= \int d^3 x \, |\bu(\bx)|^2 = \sum_{\bk} |\ukp|^2 + |\ukm|^2,\\
    H &= \int d^3 x \, \bu(\bx) \cdot \bw(\bx) = \sum_{\bk} k(|\ukp|^2 - |\ukm|^2),
\end{eqnarray} 
where $\bw(\bx)=\nabla\times\bu(\bx)$ is the vorticity.

\begin{figure*}[!htb]
\center
  \includegraphics[scale=0.25]{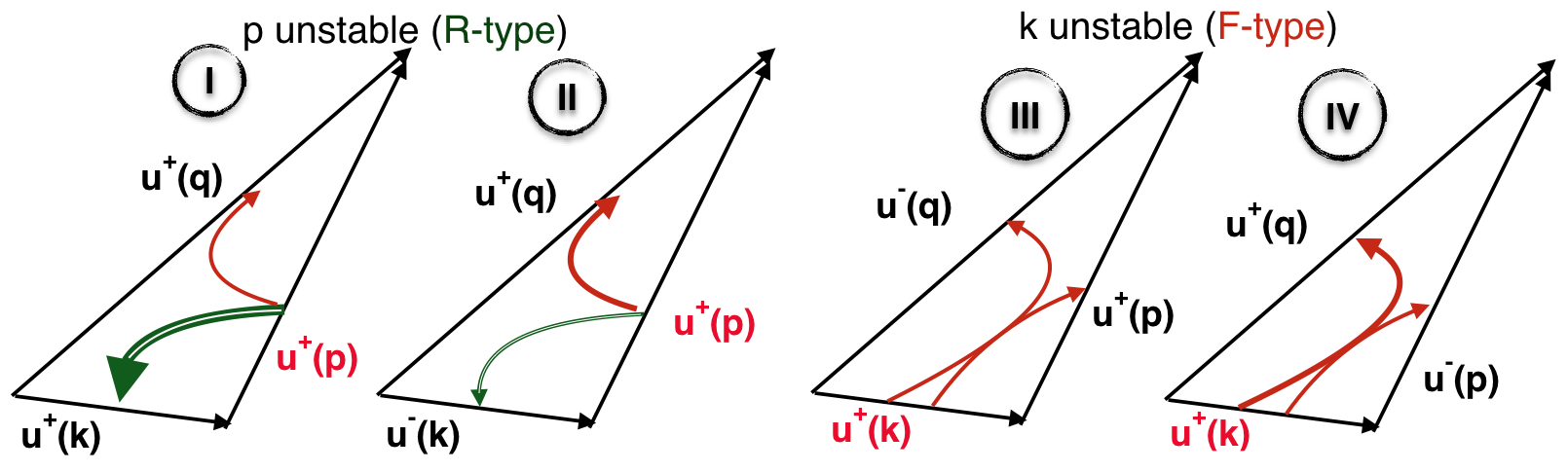}
  \caption{Schematic diagram of triadic interactions in
Navier-Stokes equations.  When Fourier-modes of velocity for two largest
wavenumbers in a triad have the same sign of helicity, there is an inverse
transfer of energy from small scales to large scales and such triads (Class-I
and Class-II) are called R-type of triads. When Fourier-modes of velocity for
two largest wavenumbers in a triad have the opposite sign of helicity, there is
forward transfer of energy from large scales to small scales and such triads
(Class-III and Class-IV) are called F-type of triads. In R-type and F-type of
triads the velocity Fourier modes with medium sized wavenumber and smallest wavenumber
are unstable, respectively, and transfer energy to other two Fourier modes. The
arrows show direction (green double-lined for inverse and red for forward) of energy
transfer. }
\label{fig:triads-all}
\end{figure*}

 It can be seen that the nonlinear term (\ref{eq:nlin}) consists of eight
possible helical combinations of the generic modes $\buk^{s_\bk}$,
$\bup^{s_\bp}$, $\buq^{s_\bq}$ forming a triad $\bk+\bp+\bq=0$ for $s_\bk=\pm$,
$s_\bp=\pm$, $s_\bq=\pm$~\cite{waleffe}. Figure \ref{fig:triads-all} shows schematic
representation of the triads which fall into four independent classes because
of the symmetry that allows simultaneous change of the sign of the helicity of
each mode.  For simplicity we assume that $k\le p \le q$. Each of these triads
conserve energy and helicity individually. 

The triads are classified as follows: Class-I contains the homochiral triads
with velocity Fourier modes having same sign of helicity for all wavenumbers,
i.e., $(\buk^+, \bup^+, \buq^+)$; Class-II contains the triads with velocity
Fourier modes having same sign of helicity for two large wavenumbers but
opposite sign of helicity for two smaller wavenumbers, i.e., $(\buk^+, \bup^-,
\buq^-)$; Class-III contains the triads with velocity Fourier modes having same
sign of helicity for the largest and the smallest  wavenumbers, i.e., $(\buk^+,
\bup^-, \buq^+)$; and Class-IV contains triads with velocity Fourier modes
having opposite sign of helicity for two larger wavenumbers but same sign of
helicity for two smaller wavenumbers, i.e., $(\buk^-, \bup^-, \buq^+)$.  

Using linear stability analysis for energy exchange among the modes of each single
triad it was argued that~\cite{waleffe} the triads of classes I do transfer
energy backward, while those of Class II,  where largest wavenumbers have same
sign of helicity, are capable of transferring energy from the unstable velocity
Fourier mode with intermediate wavenumber to the other two modes, leading to a
forward or to a backward cascade depending on the geometry of the triad
\cite{biferale2016pre,ditlevsen2016a}.  The triads in classes III and IV, where
largest wavenumbers have opposite sign of helicity, transfer energy from the
unstable velocity Fourier mode with smallest wavenumber to the other two modes
with larger wavenumbers and are responsible for forward cascade of energy.
However, in presence of more than one triads,  competing triadic interactions
do not allow simple prediction for direction of the energy transfer mechanism.
Moreover depending on the actual realization of the flow  based on the forcing
scheme, the boundary conditions, etc., different directions of the energy
transfer could be observed.  

In a turbulent flow sustained by a homogeneous and
isotropic forcing mechanism where all possible triadic interactions are present
energy is observed to be transferred forward from large to small
scales\cite{frisch}. However when the dynamics is restricted to only velocity
Fourier modes with one sign of helicity, i.e., interacting triads of Class-I
($s_\bk=s_\bp=s_\bq$), energy cascades from small scales to the large
scales~\cite{biferale2012}.  This is attributed to the fact that the second
quadratic invariant, Helicity, becomes  sign-definite for such subset of
interactions. It was also observed ~\cite{sahoo2015pre} that presence of few
percent of modes with opposite sign of helicity at all scales changes the
direction of energy transfer in a singular manner; even though triads of
classes II, III and IV are a small fraction of Class-I, they efficiently
transfer energy to the small scales. It would therefore be important to study a
system without the triads of Class-I in order to highlight their role in the
dynamics of full Navier-Stokes equations.

\begin{figure*}[!htb]
\center
  \includegraphics[scale=0.25]{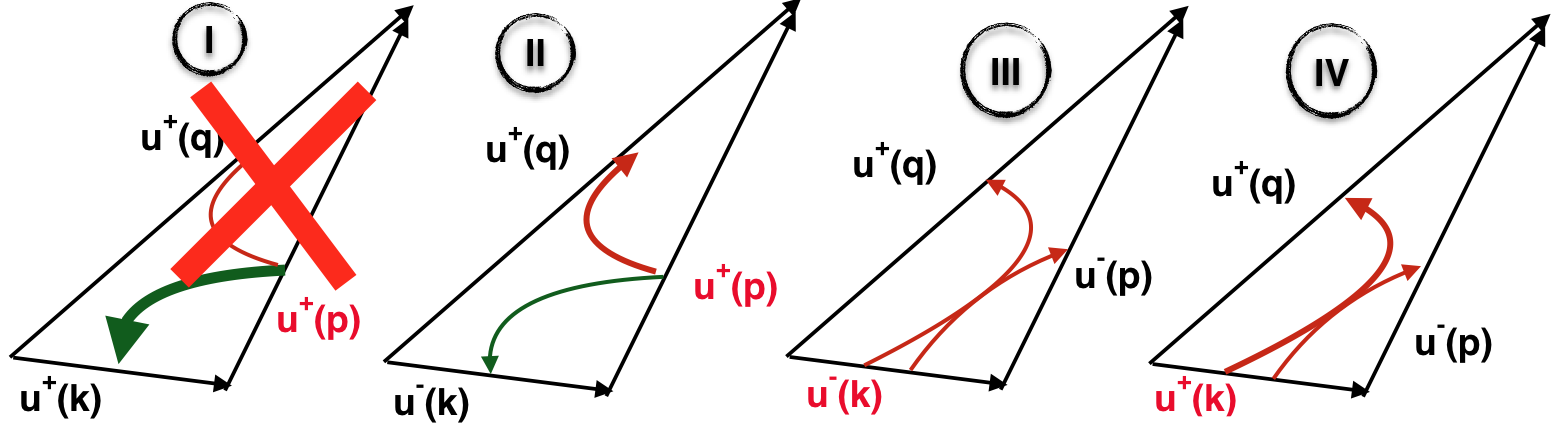}
  \caption{Schematic diagram of triadic interactions (in R2),
with no triads of Class-I, i.e., all triadic interactions involving velocity
Fourier modes of same sign of helicity were suppressed. }
\label{fig:triads-noinverse}
\end{figure*}

\begin{figure*}[!htb]
\center
  \includegraphics[scale=0.25]{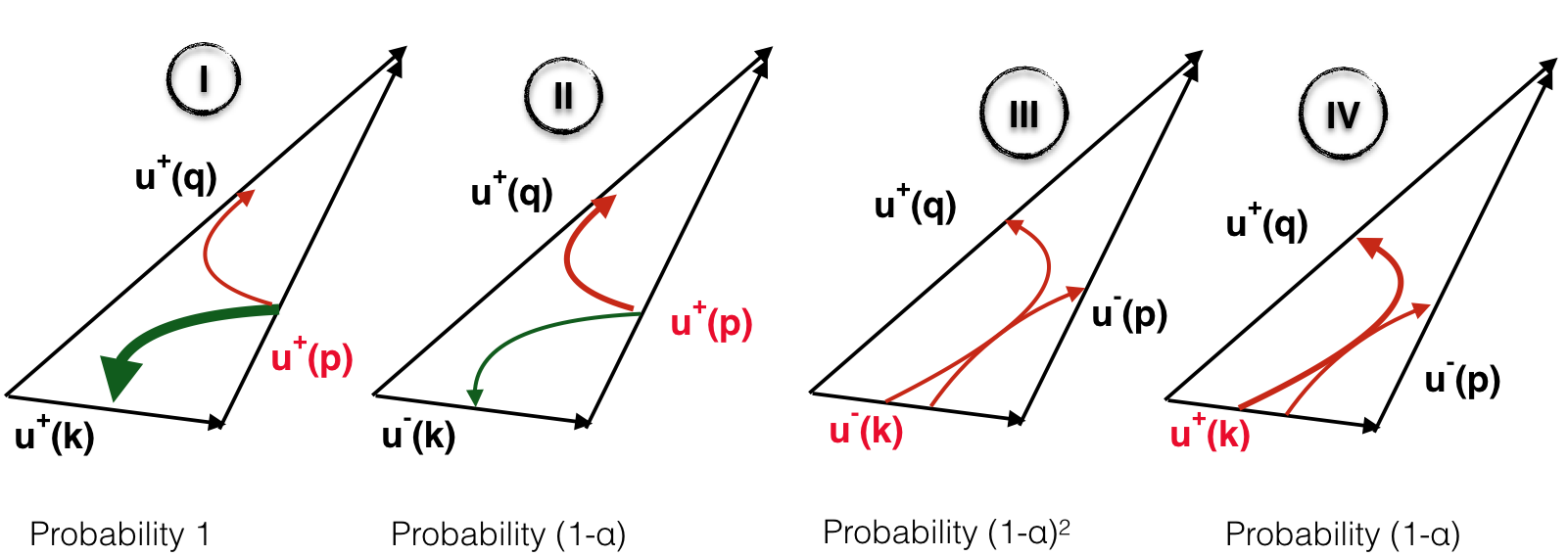}
  \caption{Schematic diagram of triadic interactions (in R3),
	  with negative Fourier-helical modes reduced with probability of $\alpha =
0.1$. } \label{fig:triads-alpha}
\end{figure*}

\section{Direct Numerical Simulations \label{sec-3}}

We have performed direct numerical simulations with a fully-dealiased,
pseudo-spectral code at resolution of $512^3$ collocation points on a triply
periodic cubic domain of size $L=2\pi$.  We used a random Gaussian forcing to
maintain a steady flow with 
\begin{equation}
\langle f_i(\bk,t) f_j(\bq,t') \rangle = F(k) \delta(\bk-\bq) \delta(t-t') Q_{i,j}(\bk),
\label{eq:force}
\end{equation}
where $Q_{ij}(\bk)$ is a projector that insures incompressibility. The
amplitude $F(k)$ is nonzero only for $ |k| \in [k_{\rm min}:k_{\rm max}]$.
This is the standard way energy is injected at large scale in simulations of turbulent flows
in order to keep homogeneity and isotropy, i.e., to keep the maximum symmetry 
in the system \cite{smith1999,biferale2012,sahoo2015pre,sahoo2017prl,borue,alvelius}. 
Table.~\ref{table1} lists the parameters of the simulations. We have used a
fully helical forcing with projection on $\hp$ in order to ensure a maximal
injection of helicity. We do not expect any dependency  of small-scale
statistics on the forcing adopted here because Navier-Stokes turbulence is
known to have universal fluctuations in the inertial and viscous ranges
\cite{schumacher2014,benzi2010} irrespective of
large-scale driving mechanism.

\begin{table}[t]
  \caption{Details of the Simulations. $N$: number of collocation points along
each axis; $L$: size of the periodic box; $\nu$: kinematic viscosity; $k_f$:
range of forced wavenumbers; $u_{\rm rms}$: rms velocity; $Re_\lambda= 
u_{\rm rms} \lambda/\nu$: Taylor-microscale Reynolds number,
 where $\lambda = \frac{2\pi}{L}\sqrt{\frac{\langle u^2(\bx) 
 \rangle}{\langle\left[\partial_x u(\bx)\right]^2\rangle}}$ is the Taylor 
 microscale; $\left<\varepsilon\right>$: mean
energy dissipation rate; $\eta$: Kolmogorov length-scale; $T_0$:
large-eddy-turnover time.}
  \label{table1}
\begin{tabular*}{\linewidth}{@{\extracolsep{\fill} } c  c  c  c  c  c  c  c  c  
c}
    \hline
    \hline
   RUN&  $N$   & $L$ & $\nu$   & $k_f$   & $u_{\rm rms}$ & $Re_\lambda$ 
   &$\left<\varepsilon\right>$ & $\eta$  & $T_0$ \\ 
    \hline
   R1 & $512$ & $2\pi$ & $0.002$ & $[1,2]$ & $3.5$      &  $220$       &    
   $3.2$     & $0.005$ & $0.3$ \\
   R2 & $512$ & $2\pi$ & $0.002$ & $[1,2]$ & $3.7$      &  $240$       &    
   $2.5$     & $0.007$ & $0.3$ \\
   R3 & $512$ & $2\pi$ & $0.002$ & $[1,2]$ & $3.5$      &  $210$       &    
   $2.8$     & $0.007$ & $0.3$ \\
    \hline
    \hline
  \end{tabular*}
\end{table}

First we carried out a simulation (R1) of standard Navier-Stokes
equations with energy injected at the large scales $k_f\in[1,2]$.  In second 
simulation (R2) we removed all the triads belonging to Class-I from the
dynamics of the Navier-Stokes equations: we solved the following modified Navier-Stokes equations
\begin{eqnarray}
\label{eq:NSa}
\partial_t \bu(\bx) &+ N[\bu(\bx),\bu(\bx)] - {\cal D}^+ N[\bu^+(\bx),\bu^+(\bx)] \\ \nonumber
&- {\cal D}^- N[\bu^-(\bx),\bu^-(\bx)] = -{\bm \nabla} p + \nu\Delta\bu(\bx)+f^+,
\end{eqnarray}
where $\nu$ is the viscosity and $p$ is the pressure. 
Such a reduction of triads preserve the conservation of energy and helicity of the system.
The triads in this simulation are shown in Fig.~\ref{fig:triads-noinverse}.
In third simulation (R3) we removed randomly $10\%$ of negatively helical velocity Fourier modes from the system using the method described in \cite{sahoo2015pre}. 
We define an operator ${D}^\alpha$ that projects each wavenumber with a probability $ 0
\le \alpha \le 1$:
\begin{equation} 
  \label{eq:projv}
  {\bu}^\alpha(\bx) \equiv D^{\alpha} {\bu}(\bx) \equiv \sum_{\bk} e^{i{\bm k}\bx}\, {\mathcal D}^\alpha_{\bk} {\buk},  
\end{equation}
where 
\({\mathcal D}^\alpha_{\bk}  \equiv (1-\gamma^\alpha_{\bk}){\mathbb I} + \gamma^\alpha_{\bk} {\mathcal P}^+_{\bk}\)
and $\gamma^\alpha_{\bk}=1$ with probability $\alpha$ or $\gamma^\alpha_{\bk}=0$ with probability $1-\alpha$. 
The $\alpha$-reduced Navier-Stokes equations ($\alpha$-NSE) are
\begin{equation} 
  \label{eq:alpha-ns}
  \partial_t \bu^\alpha = D^{\alpha}[- \bu^\alpha \cdot {\bm \nabla} \bu^\alpha -{\bm \nabla} p^\alpha] 
  +\nu \Delta \bu^\alpha, 
\end{equation}
Notice that the nonlinear terms on the {\em rhs} of (\ref{eq:alpha-ns}) are further projected by
${ D}^\alpha$ in order to enforce the dynamics  on the
selected set of modes for all times. These methods of reduction of degrees of freedom results in a loss
of Lagrangian properties of the system \cite{moffatt2014}. We chose $\alpha=0.1$ for R3. The triads present
in this simulation are shown in Fig.~\ref{fig:triads-alpha}.

\section{Results\label{sec-4}}
\begin{figure}
\center
  \includegraphics[scale=1]{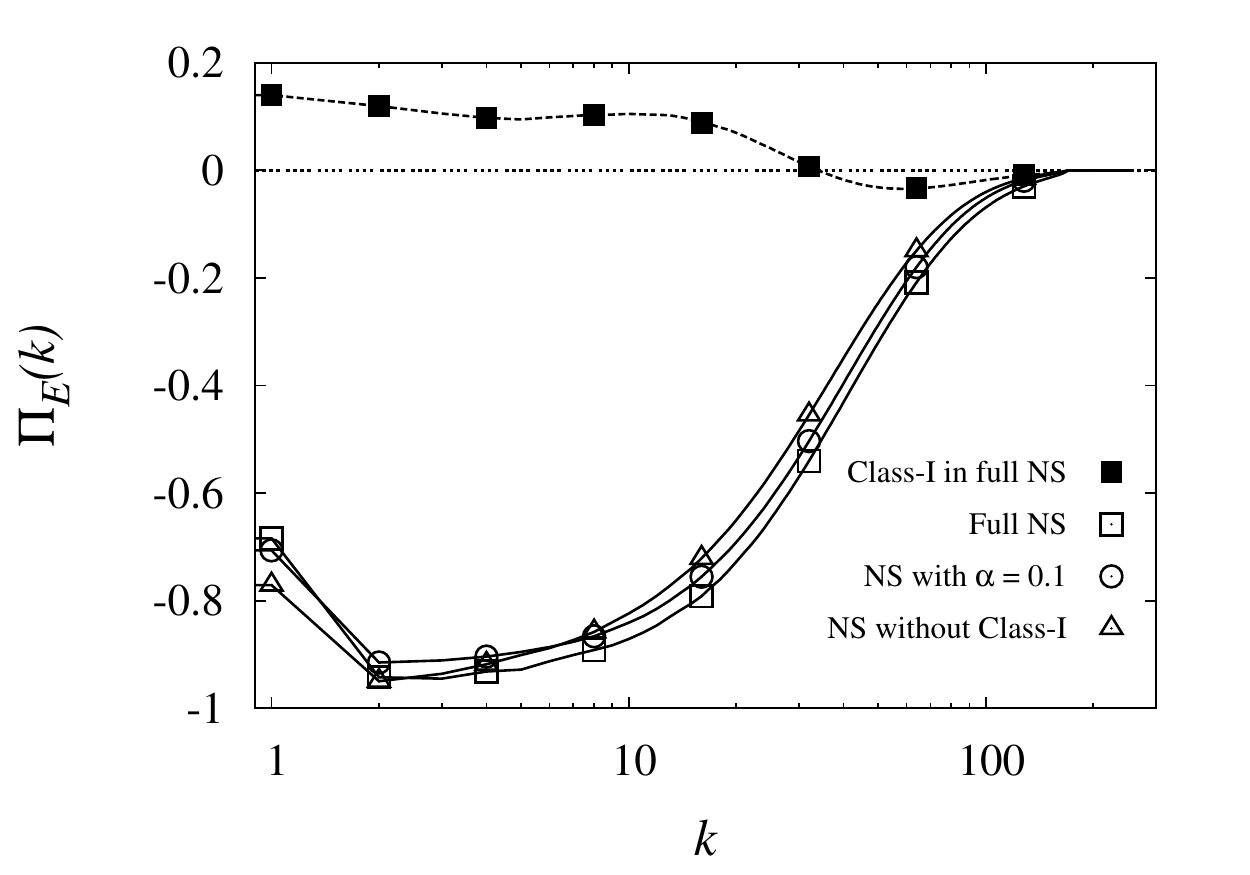}
\caption{Semi-log plots of energy fluxes. Flux of energy due to triads formed
by three Fourier modes with same sign of helicity (Class-I) in full
Navier-Stokes equations (R1) are shown by filled squares whereas the total flux of
energy is shown by empty squares. Total energy flux from simulation (R3) with 
$\alpha=0.1$ are shown by circles whereas triangles show the same from
simulation (R2) of Navier-Stokes equations without triads of Class-I.}
\label{fig:fluxes}
\end{figure}

	We measured the energy flux due to the nonlinear terms, given in Eq.~(\ref{eq:nlin}), 
\begin{equation}
{\rm \Pi}_E(k)=\sum_{|\bk'|<k} {\hat {\bm u}}_{\bm k'}^* \cdot \hat\bN_{\bk'},  
\label{eq:flux}
\end{equation}
across a wavenumber $k$, for all three cases. 
We show the total energy flux and the energy
flux due to only homochiral triads (Class-I), by using either of the projected 
velocity modes $\bukpm$ in Eqs.~(\ref{eq:nlin}) and (\ref{eq:flux}) 
in the full Navier-Stokes equations
(no mode reduction) in Fig.~\ref{fig:fluxes}. 
The flux due to triads of Class-I  has opposite sign to
that of total flux indicating that those interactions contribute with an
inverse transfer of energy already in the full equations as observed in Ref.~\cite{alexakis}. Also in
Fig.~\ref{fig:fluxes} we compare the total energy flux in full Navier-Stokes
(from simulation R1), in $\alpha$-reduced Navier-Stokes (R3) and in
Navier-Stokes without the triads of Class-I (from simulation R2); there is no
significant difference in the total flux. This is due to the fact
that the net total flux has strictly zero backward energy
transfer for all the three cases\footnote{ In ~\cite{sahoo2015pre} it was shown that the energy transfer is reversed only when
  almost all negative modes are removed, i.e.,   $\alpha \sim 1$. }.

\begin{figure}
\center
  \includegraphics[scale=1]{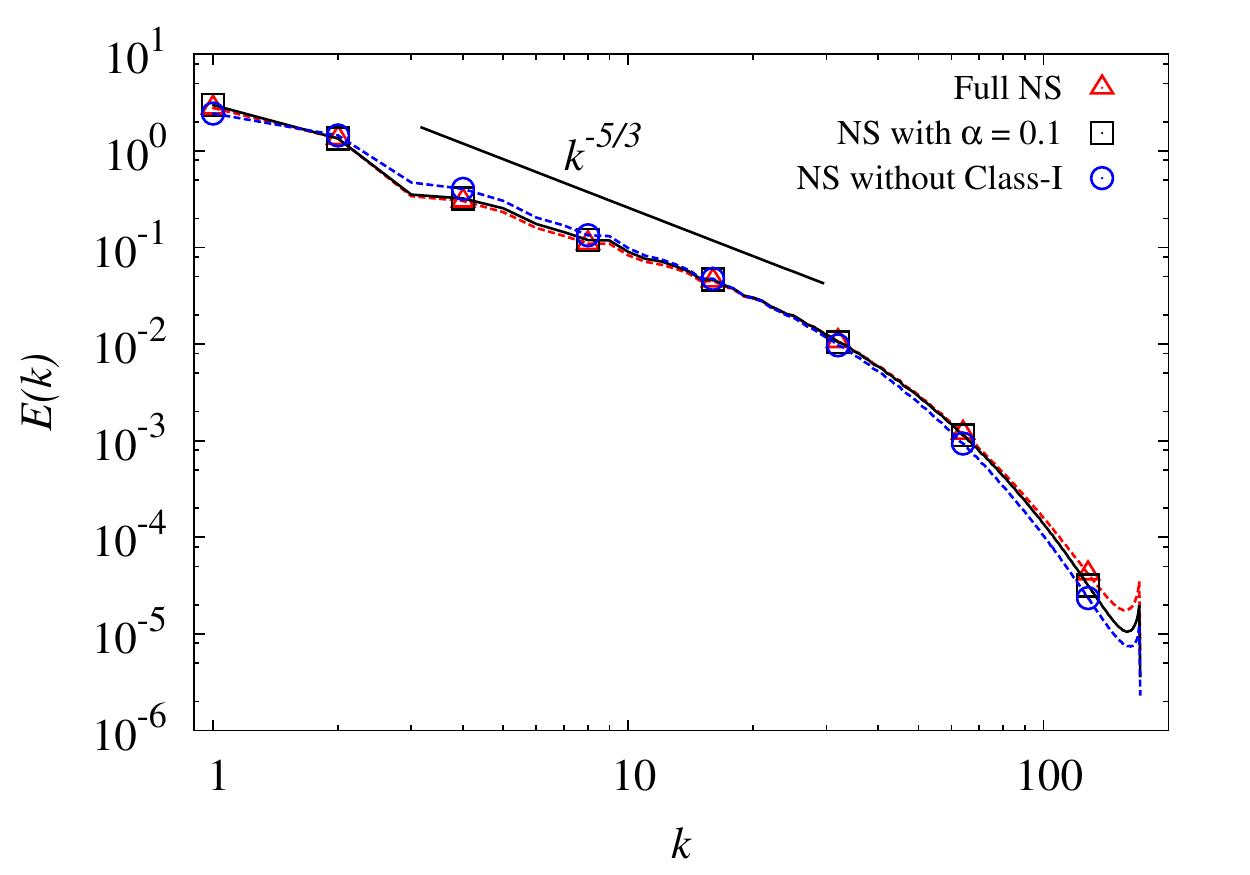}
\caption{Log-log plots of energy spectra. Triangles, squares and circles show
the spectra of energy from simulations of full Navier-Stokes equations (R1),
$\alpha$-reduced Navier-Stokes equations (R3) and Navier-Stokes equations without
Class-I triads (R2), respectively. The black line shows $k^{-5/3}$ scaling for reference.}
\label{fig:spectra}
\end{figure}

In Fig.~\ref{fig:spectra} we compare the energy spectra for the same three cases (R1, R2 and R3) which are
indistinguishable from each other. Energy spectra are not sensitive to reduction of triads as long as 
triads facilitating forward energy cascade are  present in the system. This is in agreement with
earlier observation ~\cite{sahoo2015pre}.

\begin{figure}
\center
  (a)\includegraphics[scale=1]{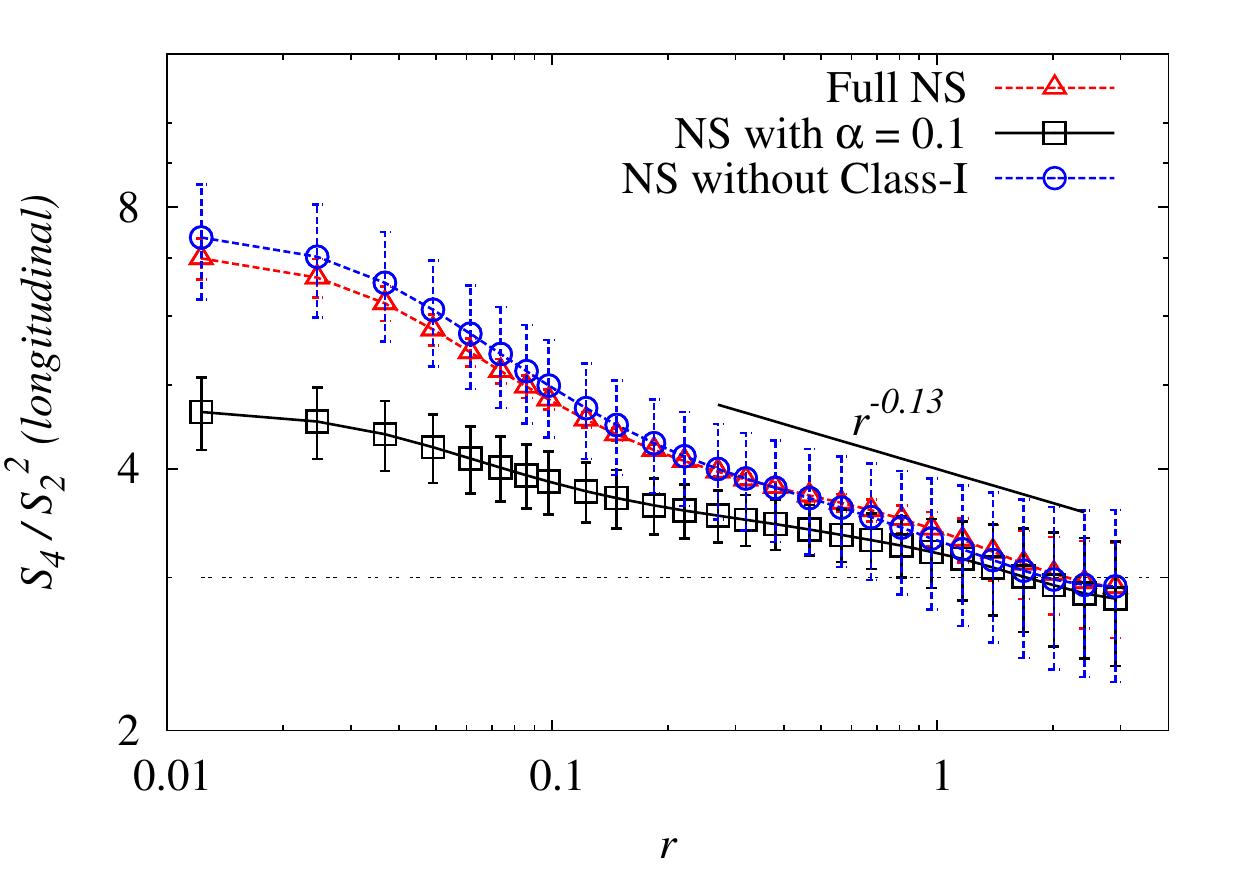}
  (b)\includegraphics[scale=1]{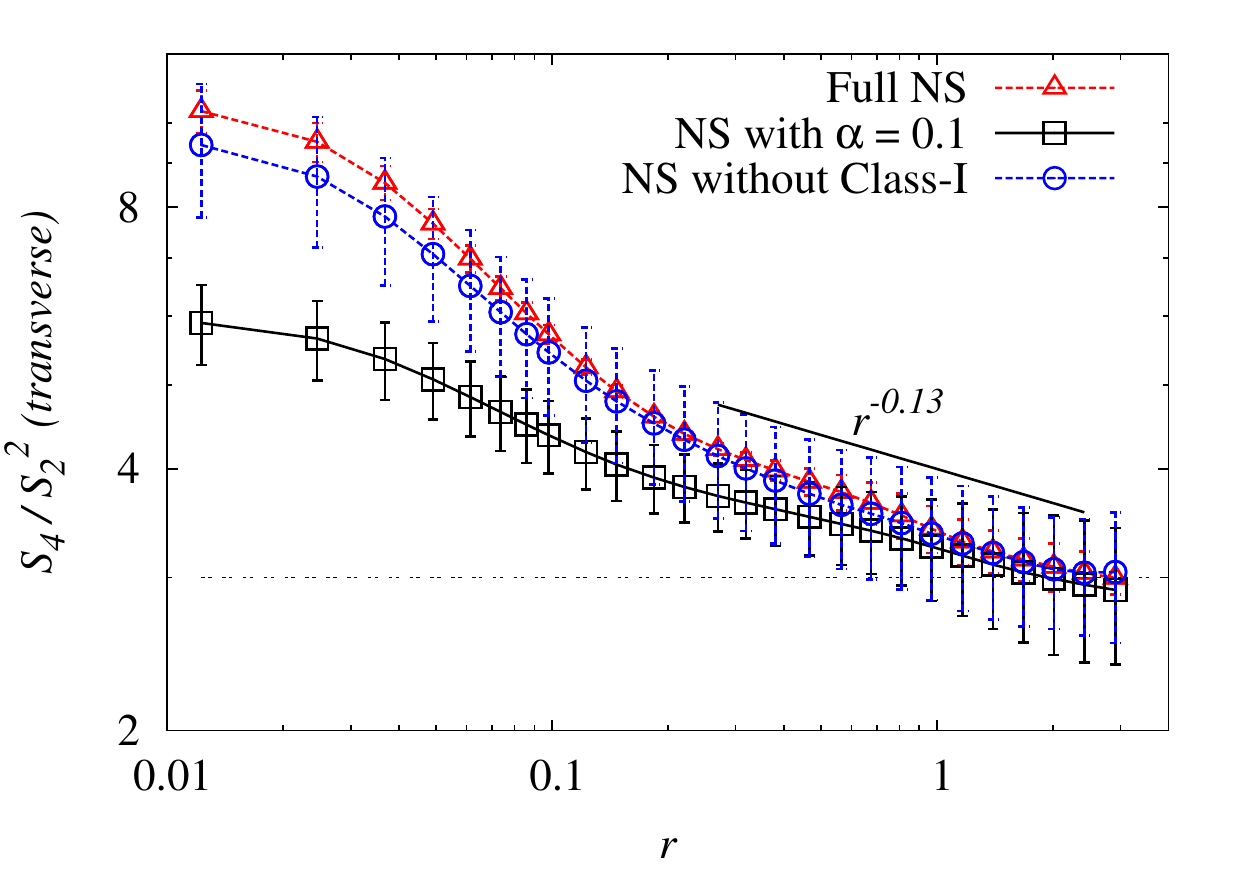}
  \caption{(a) Flatness of longitudinal velocity increments from simulations of full
		  Navier-Stokes equations (R1), $\alpha$-reduced Navier-Stokes equations (R3)
and Navier-Stokes equations without Class-I triads (R2) are shown by triangles,
squares and circles, respectively. (b) Flatness of transverse velocity increments. Error-bars show
the fluctuations in the steady-state.}
\label{fig:flatness}
\end{figure}

One  goal of this paper is to study the effects of removing 
Class-I triads on the intermittency of the system. We measured the flatness, defined as
\begin{equation}
F(r)= \frac{S_4(r)}{[S_2(r)]^2},
\end{equation}
and hyperflatness, defined as
\begin{equation}
H(r)= \frac{S_6(r)}{[S_2(r)]^3},
\end{equation}
of longitudinal velocity increments 
\begin{equation}
\delta_{r} u_L = [\bu(\bx+\br) - \bu(\bx)]\cdot\frac{\br}{r},
\end{equation}
and transverse velocity increments
\begin{equation}
\delta_{r} u_T = [\bu(\bx+\br) - \bu(\bx)]\cdot\frac{\bf r'}{r'},
\end{equation}
where ${\bf r'}$ is perpendicular to the direction of $\br$, shown in Fig.~\ref{fig:flatness} for three cases R1-R3. 
Structure functions, longitudinal and transverse, of order $p$ are defined as
\begin{equation}
	S_p^{L,T}(r) = \left<(\delta_{r} u_{L,T})^p\right>,
\end{equation}
where angular brackets denote spatial average. 

\begin{figure}
\center
  (a)\includegraphics[scale=1]{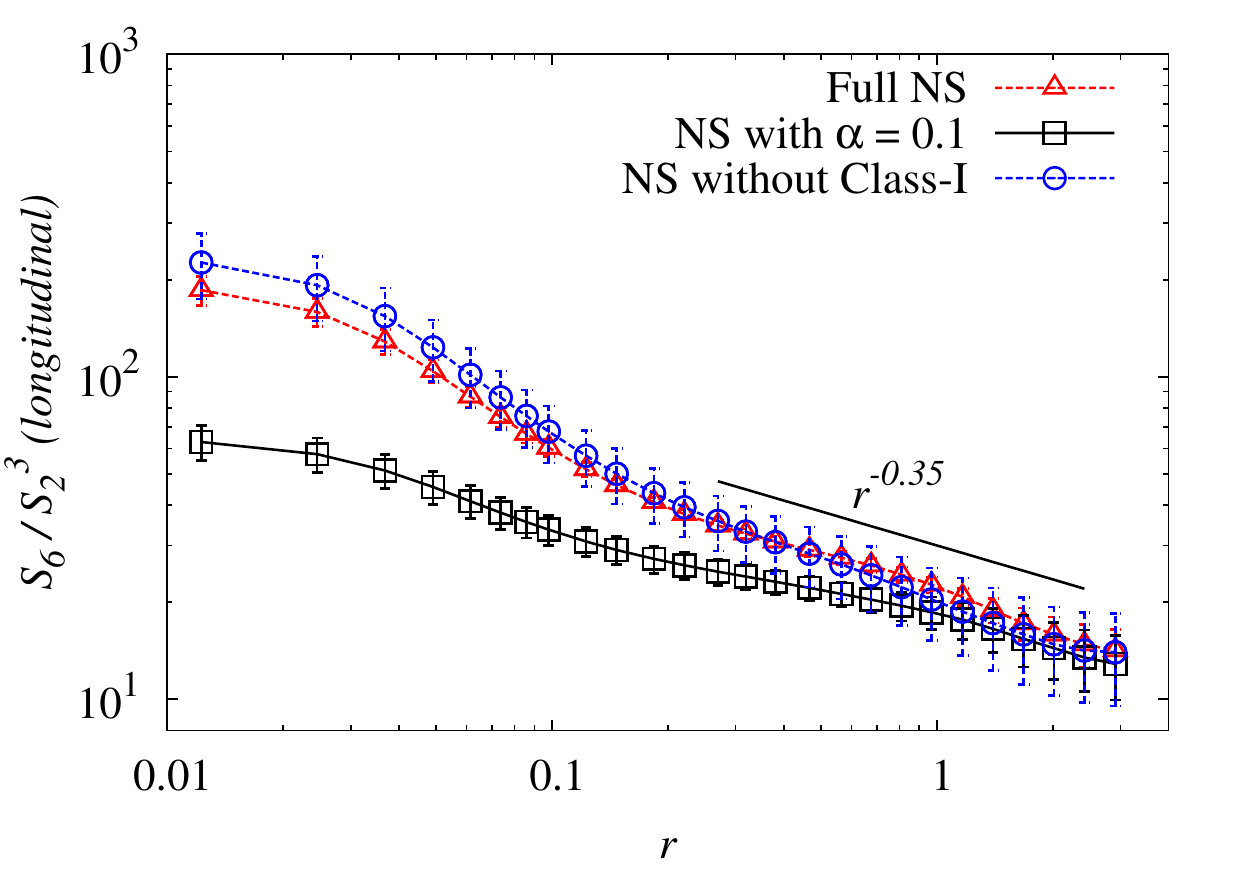}
  (b)\includegraphics[scale=1]{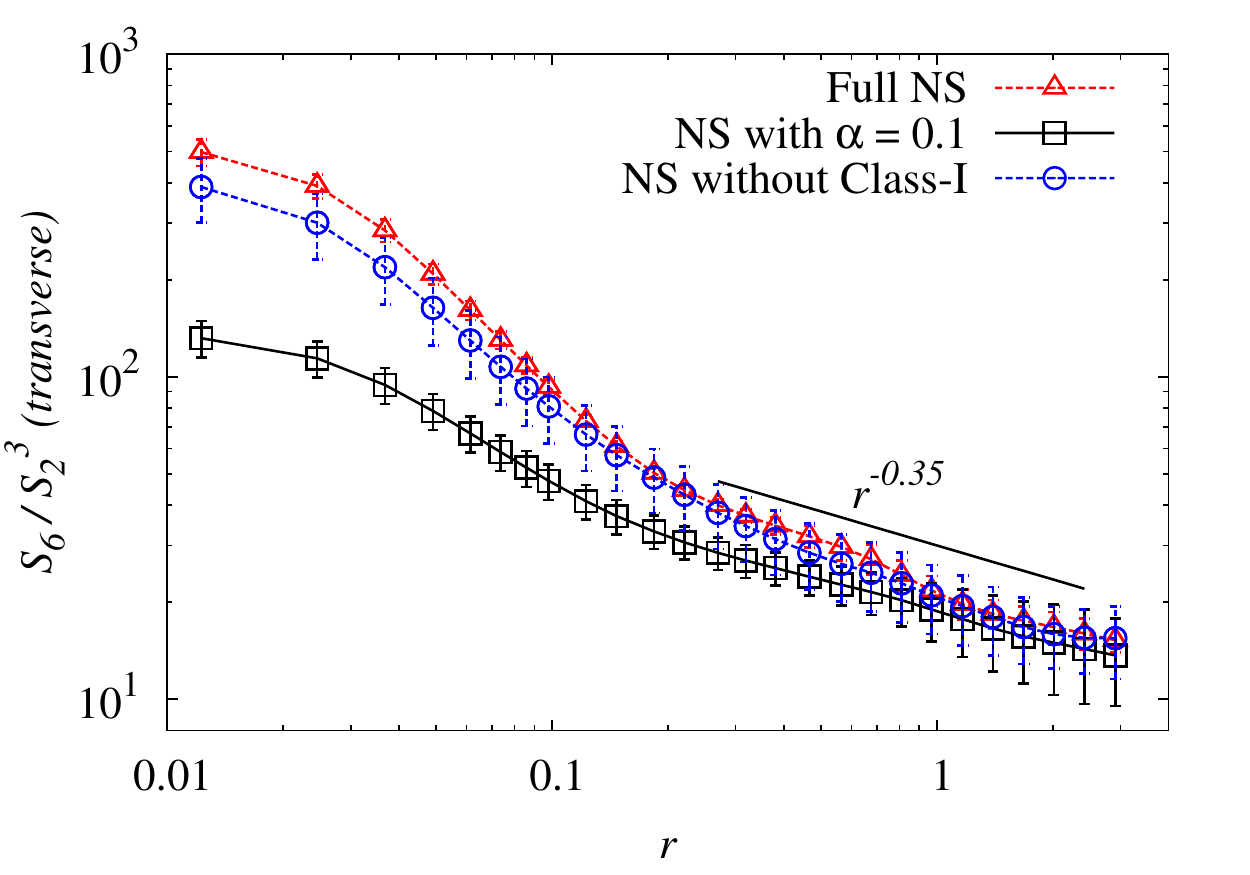}
  \caption{(a) Hyperflatness of longitudinal velocity increments from simulations of full
		  Navier-Stokes equations (R1), $\alpha$-reduced Navier-Stokes equations (R3)
and Navier-Stokes equations without Class-I triads (R2) are shown by triangles,
squares and circles, respectively. (b) Hyperflatness of transverse velocity increments. Error-bars show
the fluctuations in the steady-state.}
\label{fig:hyperflatness}
\end{figure}

\begin{figure}
\center
  (a)\includegraphics[scale=1]{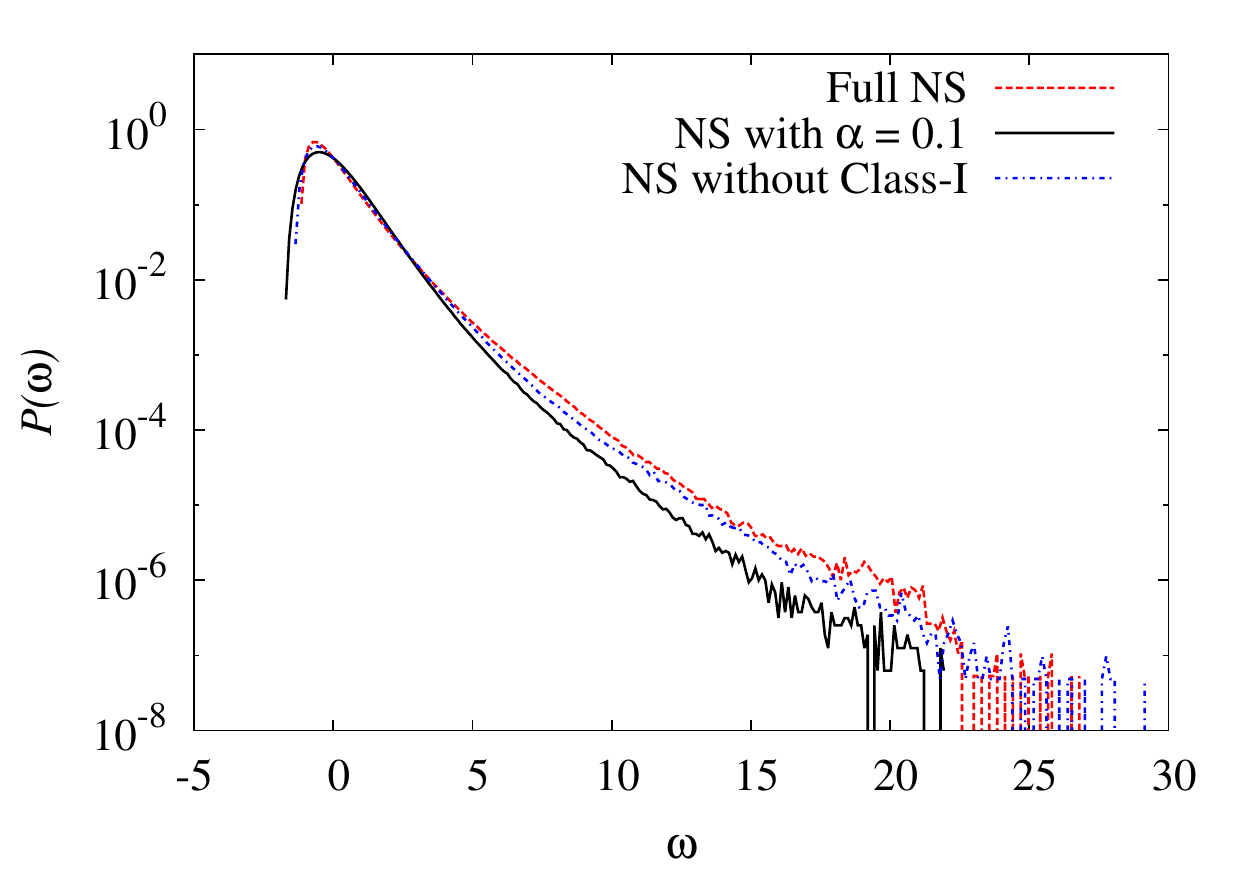}
  (b)\includegraphics[scale=1]{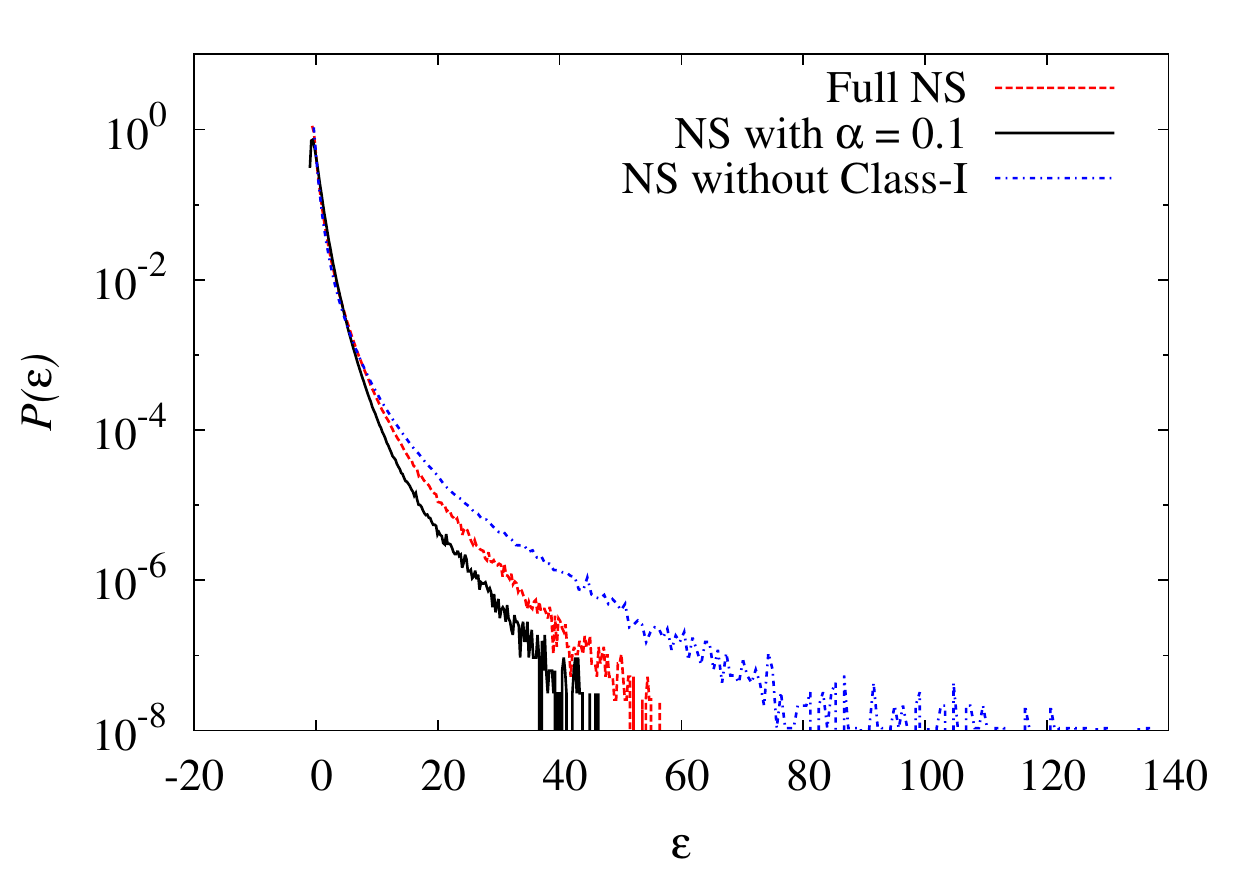}
  \caption{(a) Plots of probability distribution function (pdf) of $\omega$, where  
	  $\omega^2 = |\nabla \times \bu|^2$ is the enstrophy, from simulations of full
	  Navier-Stokes equations (R1), $\alpha$-reduced Navier-Stokes equations (R3)
and Navier-Stokes equations without Class-I triads (R2) are shown by triangles,
squares and circles, respectively. (b) Plots of pdf of local energy dissipation rates
$\varepsilon = 2\nu(\sum_{i,j}\partial_iu_j - \partial_ju_i)^2$.}
\label{fig:pdf}
\end{figure}

The Flatness of both longitudinal and transverse velocity
increments in the inertial range, for the full Navier-Stokes (R1) and the Navier-Stokes with only heterochiral
triads (R2) are comparable within the error-bars (See
Fig.~\ref{fig:flatness}a). This indicates that homochiral (Class-I)
triads, which are responsible for inverse energy transfer, have no significant
role in intermittency in the inertial range of scales. However the flatness
for the $\alpha$-reduced Navier-Stokes equations (R3) is much lower than the
full Navier-Stokes equations (R1). A similar behaviour is also observed for
the longitudinal and transverse hyper-flatness which are shown in
Fig.~\ref{fig:hyperflatness}.  It is also observed that in absence of
homochiral triads (R2) the intensity of the  longitudinal ﬂatness at the
gradient scale  is marginally higher than the full
Navier-Stokes case (R1) and the opposite is measured for the transverse
increments, which could be an indication that the presence/absence of
homochiral triads slightly modifies the small-scale vortical structure of the
flow.  

To further investigate this aspect, we measured the probability
distribution function (pdf) of local energy dissipation rate and of local
enstrophy (see Fig.~\ref{fig:pdf}). As inferred from the measurement of the ﬂatness we observed
that the pdf of energy dissipation  has a longer tail for the case of only
heterochiral triads and the opposite happens for the enstrophy distribution,
confirming that the absence of homochiral triads might have a different impact in regions of high strain or high rotation. 

\begin{figure}
\center
  \includegraphics[scale=0.35]{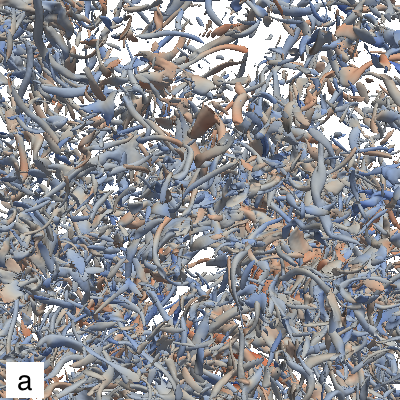}
  \includegraphics[scale=0.35]{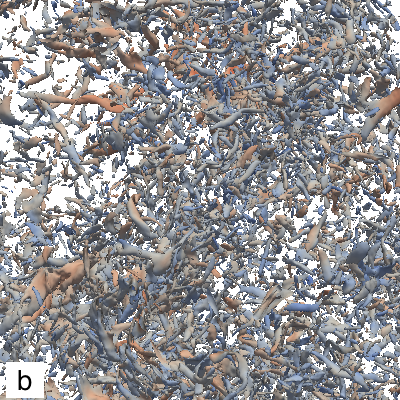}
  \includegraphics[scale=0.35]{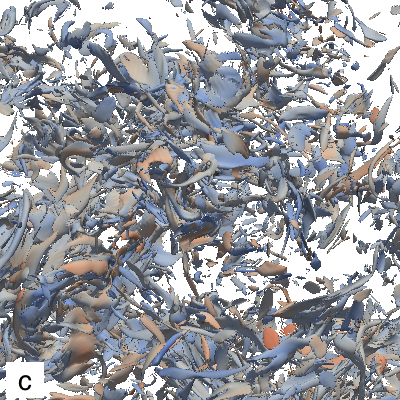}
  \caption{Isovorticity surfaces for (a) Full Navier-Stokes (R1), (b)
	  $\alpha$-reduced Navier-Stokes with $\alpha=0.1$ (R3) and (c) Navier-Stokes without the
	  Class-I triads (R2).  The isovalues of the vorticity is three standard deviations
above the mean value. The color code correspond to the value of helicity: from
high positive values (red) to high negative values (blue).}
\label{fig:isosurfaces}
\end{figure}

Visualisation of the flow field with plots of isovorticity surfaces are shown in Fig.~\ref{fig:isosurfaces}.
In both cases where we have removed the Class-I triads or a fraction of triads
of classes other than Class-I the filament-like structures are reduced. In
absence of Class-I triads we observe more sheet-like structures. The visible
change in the `coherency' of the flow together with the intermittency
robustness as a function of the removal of Class-I triads suggest that the main
signature of anomalous scaling in the inertial range of the full Navier-Stokes equations is not
strongly connected to any clearly detectable `coherent structure'.

\section{Conclusions\label{sec-5}}

We carried out direct numerical simulations of 3D
Navier-Stokes equations and of the Navier-Stokes equations without the
triads formed by homochiral velocity Fourier modes (here taken the
positive ones). We observed that inertial range  intermittency remains
almost unaffected confirming that the forward energy cascade is mainly
dominated by heterochiral triads. On the other hand, by removing
negative helical modes also on the heterochiral triads, intermittency
strongly reduces, suggesting that the formation of small-scale intense
events needs almost all  triads that transfer energy forward.
A small change at the scale crossing between viscous
and inertial terms is  observed when homochiral triads are removed from the
dynamics in agreement with the presence of more sheet-like structures in the
flow. It would be interesting to extend this kind of studies to less
symmetric flow configuration. In particular it might be key to apply it
for the case of turbulence under rotation, where previous works have
shown that helicity plays a role in enhancing the inverse energy
cascade regime \cite{pouquet2010}. On the other hand, rotating turbulence tends to become
quasi-2D  in the limit of  very intense rotation rate \cite{biferale2016prx,sagar2007epl} , and therefore  there must exists a trade-off between the role
played by helicity (exactly vanishing in 2D) and  rotation \cite{bbl2017}. Similarly,
it is not known the role played by homochiral and heterochiral triads
in strongly anisotropic flows as for the case of homogeneous shear and for
decaying turbulence. Work in this direction is ongoing and it will be reported
elsewhere.

\section{Acknowledgement}

We acknowledge funding from the European Research Council under the European
Union's Seventh Framework Programme, ERC Grant Agreement No 339032 and support
from COST Action MP1305.  GS acknowledges support from Atmospheric Mathematics
(AtMath) collaboration at University of Helsinki.

\end{document}